\documentclass[12pt,epsf]{article}
\usepackage{graphicx}

\newcommand{\beq}{\begin{equation}}
\newcommand{\eeq}{\end{equation}}
\newcommand{\bea}{\begin{eqnarray}}
\newcommand{\eea}{\end{eqnarray}}

\begin{document}

\begin{titlepage}
\vspace{1.0cm}
\begin{center}
{\Large\bf Towards A Realistic Grand Gauge-Higgs Unification Scenario\footnote{
Talk given at International Workshop on Grand Unified Theories : 
Current Status and Future Prospects (GUT07), December 17-19 2007, Kusatsu, Japan. 
This talk is based on a paper collaboration with C.S. Lim \cite{LM2}.}} 
\end{center}
\vspace{15mm}

\begin{center}
{\Large Nobuhito Maru}
\end{center}
\vspace{0.5cm}
\centerline{{\it Department of Physics, Kobe University,
Kobe 657-8501, Japan}}
\vspace{1cm}

\begin{center}
{\bf abstract}
\end{center}
In this talk, we discuss an attempt to construct a realistic model 
of the grand gauge-Higgs unification. 
We investigate a 5D $SU(6)$ grand gauge-Higgs unification model 
compactified on an orbifold $S^1/Z_2$. 
Ordinary quarks and leptons, together with right-handed neutrinos, 
are just accommodated into a minimal set of representations of the gauge group, 
without introducing any exotic states in the same representations. 
The proton decay turns out to be forbidden at least at the tree level. 
We also find a correct electroweak symmetry breaking 
$SU(2)_L \times U(1)_Y \to U(1)_{em}$ is easily realized 
by introducing suitable number of adjoint fermions. 

\end{titlepage}
%\maketitle

%%%%%%%%%%%%%%%%%%%%%%%%%%%%%%%%%%%%%%%%%%%%
%% MAINMATTER
%%%%%%%%%%%%%%%%%%%%%%%%%%%%%%%%%%%%%%%%%%%%

The hierarchy problem, especially the problem of 
how to stabilize the Higgs mass under the quantum correction, 
has played a key role to motivate the physics beyond the standard model. 
Almost all possible scenarios to solve the problem invoke 
to some sort of symmetry in order to protect the Higgs mass 
at the quantum level. 
Supersymmetry is the most popular scenario and has been extensively discussed. 

Recently the gauge-Higgs unification scenario \cite{Manton, Fairlie, Hosotani} 
has obtained a revived interest as a possible new avenue 
to solve the problem \cite{HIL}. 
In this scenario the Higgs is regarded as the extra space component of 
higher dimensional gauge fields and the quantum corrections to the Higgs mass becomes finite 
by higher dimensional gauge  symmetry without relying on the supersymmetry. 
This fact has been verified \cite{HIL}-\cite{LMH} at one loop level and \cite{MY} at two loop level 
in various models by explicit calculations. 
By the same reasoning, the gauge-Higgs unification predicts the physical observables to be finite 
thanks to the higher dimensional gauge invariance in spite of its nonrenormalizability. 
The divergence structure of S and T parameters in the gauge-Higgs unification was investigated 
and an interesting relation between them, which is only spcific to the gauge-Higgs unification, 
was predicted to be finte in six dimensional case \cite{LM}.
The anomalous magnetic moment in the gauge-Higgs unification 
was found to be finite in arbitrary dimension \cite{ALM}.
Recently, the gluon fusion process \cite{Falkowski, MO, LR} 
and the two photon decay of the Higgs boson \cite{MO} crucial for the LHC physics 
were studied and a very interesting prediction was reported. 
There have been many works on rich structure of the theory and its phenomenology, 
see \cite{KLY}-\cite{ST}. 

Another possible interesting scenario is to regard the Higgs 
as a pseudo Nambu-Goldstone (PNG) boson 
due to the breakdown of some global symmetry. 
As far as the global symmetry is larger than local gauge symmetry, 
even though some N-G bosons are absorbed to gauge bosons 
via Higgs mechanism on the spontaneous symmetry breaking, 
there should remain some physical PNG bosons, 
which can be identified as the Higgs bosons \cite{IKT}-\cite{Berezhiani}. 
The scenario faces a difficulty at quantum level, 
once gauge interactions are switched on. 
Namely, the Higgs mass suffers from a quadratic divergence, 
essentially because the original global symmetry is partly gauged 
and therefore the global symmetry is hardly broken by the gauge couplings. 
However, such difficulty may be avoided, 
once direct products of identical global symmetries are taken. 
It is interesting to note that such ``dimensional deconstruction"  
or related ``little Higgs" scenarios \cite{ACG} can be regarded as 
a kind of gauge-Higgs unification, 
where the extra space has finite number of lattice points. 
In fact, it is a recent remarkable progress to have established 
the relation between a four dimensional theory with global symmetry $G$ 
and a five dimensional gauge theory with gauge symmetry $G$, 
through the AdS/CFT correspondence (holographic approach) \cite{holography, holography2}. 

The gauge hierarchy problem was originally discussed 
in the framework of Grand Unified Theory (GUT) 
as the problem to keep the discrepancy between the GUT scale 
and the weak scale. 
So it will be meaningful to test the possible scenarios 
in the framework of GUT. 

The PNG boson scenario in the framework of GUT was discussed 
long time ago \cite{IKT}. 
Since the global symmetry needs to be larger than the gauge symmetry $SU(5)$, 
a minimal model with an $SU(6)$ global symmetry was proposed. 
As explained above, 
unfortunately the Higgs boson suffers from a quadratic divergence 
at quantum level, and SUSY was introduced to eliminate 
the unwanted divergence.  

In this work, in view of the recent progress mentioned above, 
we attempt to construct a GUT model based on the scenario of 
gauge-Higgs unification, which is refered to as ``grand gauge-Higgs unification". 
A nice thing in our scenario is that the Higgs mass is automatically 
stabilized at the quantum level without relying on the SUSY.  

What we adopt is a minimal grand gauge-Higgs unification model, 
i.e. 5-dimensional (5D) GUT with an $SU(6)$ gauge symmetry. 
It is interesting to note that an $SU(6)$ symmetry emerges again, 
as suggested by the AdS/CFT correspondence. 
This is because in the gauge-Higgs unification, 
the gauge symmetry needs to be enlarged from the minimal one $SU(5)$, 
since Higgs inevitably belongs to the adjoint representation of 
the gauge group while the Higgs should behave 
as the fundamental representation of $SU(5)$. 

In addition to the finite Higgs mass, as a bonus, 
we find that the sector of fermionic zero-mode of the theory 
just accommodates three generations of quarks and leptons. 
Namely, we do not encounter the problem of introducing massless exotic particles 
in the representations quarks and leptons belong to, 
which often happens in the gauge-Higgs unification.
Another remarkable feature, we will see below, 
is that the dangerous proton decay due to the exchange of GUT particles 
turns out to be prohibited at least at the tree level 
without any symmetry. 
This is due to the splitting multiplet mechanism. 

We will also discuss how the desirable gauge symmetry breaking is realized 
via Hosotani mechanism \cite{Hosotani}. 
We will see the desirable pattern of gauge symmetry breaking 
is realized without introducing additional scalar matter fields. 

In ref. \cite{BN}, 
an elegant 5D $SU(6)$ grand gauge-Higgs unification model was discussed, 
but as a SUSY theory. 
In this context, a non-SUSY $SU(6)$ grand gauge-Higgs unification model 
has been already studied \cite{HHKY}, 
where the main focus was in the pattern of electroweak symmetry breaking  
and a viable Higgs mass satisfying the experimental lower limit was obtained. 
   
The set-up of our model concerning gauge-Higgs sector just 
follows the model of \cite{HHKY} and \cite{BN}. 
%as follows. 
The 5D space-time we consider has an extra space compactified 
on an orbifold $S^1/Z_2$ with a radius $R$, whose coordinate is $y$.  
On the fixed points $y=0, \pi R$, 
the different $Z_2$ parities are assigned as 
\bea
P = {\rm diag}(+,+,+,+,+,-)~{\rm at}~y=0, \quad
P'= {\rm diag}(+,+,-,-,-,-)~{\rm at}~y=\pi R,
\label{parity}
\eea
which implies the gauge symmetry breaking pattern 
\bea
&&SU(6) \to SU(5)\times U(1)~{\rm at}~y=0, \\
&&SU(6) \to SU(2) \times SU(4) \times U(1)~{\rm at}~y = \pi R. 
\eea
in each fixed point. 
These symmetry breaking patterns are inspired by \cite{HHKY, IKT, BN}. 
It is instructive to see the concrete parity assignments 
of each component of the 4D gauge field $A_{\mu}$ and 4D scalar field 
$A_5$,   
\bea
A_\mu &=& 
\left(
\begin{array}{cccccc}
(+,+) & (+,+) & (+,-) & (+,-) & (+,-) & (-,-) \\
(+,+) & (+,+) & (+,-) & (+,-) & (+,-) & (-,-) \\
(+,-) & (+,-) & (+,+) & (+,+) & (+,+) & (-,+) \\
(+,-) & (+,-) & (+,+) & (+,+) & (+,+) & (-,+) \\
(+,-) & (+,-) & (+,+) & (+,+) & (+,+) & (-,+) \\
(-,-) & (-,-) & (-,+) & (-,+) & (-,+) & (+,+) \\
\end{array}
\right), \\
A_5 &=& 
\left(
\begin{array}{cccccc}
(-,-) & (-,-) & (-,+) & (-,+) & (-,+) & (+,+) \\
(-,-) & (-,-) & (-,+) & (-,+) & (-,+) & (+,+) \\
(-,+) & (-,+) & (-,-) & (-,-) & (-,-) & (+,-) \\
(-,+) & (-,+) & (-,-) & (-,-) & (-,-) & (+,-) \\
(-,+) & (-,+) & (-,-) & (-,-) & (-,-) & (+,-) \\
(+,+) & (+,+) & (+,-) & (+,-) & (+,-) & (-,-) \\
\end{array}
\right),  
\eea
where $(+,-)$ means that the $Z_2$ parity is even (odd) at $y=0(y=\pi R)$, 
for instance.   
Kaluza-Klein (KK) mode expansion in each type of the parity assignment 
is given by 
\bea
\Phi_{(+,+)}(x,y) &=& \frac{1}{\sqrt{2\pi R}}
\left[
\phi^{(0)}_{(+,+)}(x) +\sqrt{2}\sum_{n=1}^\infty \phi_{(+,+)}^{(n)}(x)
\cos \left(\frac{n}{R}y \right) \right], 
\label{++} \\
\Phi_{(+,-)}(x,y) &=& \frac{1}{\sqrt{\pi R}}
\sum_{n=0}^\infty \phi_{(+,-)}^{(n)}(x)
\cos \left(\frac{n+\frac{1}{2}}{R}y \right), 
\label{+-} \\
\Phi_{(-,+)}(x,y) &=& \frac{1}{\sqrt{\pi R}}
\sum_{n=0}^\infty \phi_{(-,+)}^{(n)}(x)
\sin \left(\frac{n+\frac{1}{2}}{R}y \right), 
\label{-+} \\
\Phi_{(-,-)}(x,y) &=& \frac{1}{\sqrt{\pi R}}
\sum_{n=1}^\infty \phi_{(-,-)}^{(n)}(x)
\sin \left(\frac{n}{R}y \right). 
\label{--}
\eea
Noting that the 4D massless KK zero mode appears only in the $(+,+)$ component, 
the gauge symmetry breaking by orbifolding is found 
(see $A_\mu$ parity assignment) to be 
\bea
SU(6) \to SU(3)_C \times SU(2)_L \times U(1)_Y \times U(1)_X. 
\eea
Here the hypercharge $U(1)_Y$ is contained in the upper-left 
$5 \times 5$ block of Georgi-Glashow $SU(5)$. 
Therefore, we obtain 
\bea
g_3 = g_2 = \sqrt{\frac{5}{3}}g_Y,  
\eea
at the unification scale, which will be not far from $1/R$. 
This means that the Weinberg angle is just the same as the Georgi-Glashow 
$SU(5)$ GUT, namely $\sin^2 \theta_W = 3/8$ ($\theta_W$: Weinberg angle) 
at the classical level. 
In fact, we can explicitly confirm it for a ${\bf 6^*}$ representation given 
in (\ref{12}) below, 
\bea
\sin^2 \theta_W = \frac{{\rm Tr}I_3^2}{{\rm Tr}Q^2} 
= \frac{(\frac{1}{2})^2+(-\frac{1}{2})^2}{(\frac{1}{3})^2 \times 3 + (-1)^2} 
=\frac{3}{8} 
\eea
where $I_3$ is the third component of $SU(2)_L$ 
and $Q$ is an electric charge. 
In order to compare with the experimental data, 
the gauge coupling running effects have to be taken into account. 
Such a study is beyond the scope of this work. 
Since Higgs belongs to the doublet of $SU(2)_L$ and 
the electroweak gauge symmetry is embedded into ordinary $SU(5)$, 
the Z boson mass is given as
\bea
M_Z^2 = \frac{M_W^2}{\cos^2 \theta_W} 
= \sqrt{\frac{8}{5}}M_W \simeq 102 {\rm GeV}
\eea
at the classical level.

On the other hand, concerning $A_5$, 
the zero mode appears only in the doublet component, as we wish. 
The colored Higgs has a mass at least of the order of $1/R$, 
and the doublet-triplet splitting is automatically realized \cite{Kawamura}.  

%%%%%%%%%%%%%%%%%%%%%%%%%%%%%%%%%%%%%%%%%%%%%%%%%%%%%%%%%%%%%%%%%%%%%%
Let us now turn to the non-trivial question of 
how quarks and leptons can be accommodated into the representations of $SU(6)$, 
without introducing exotic states in the zero mode sector. 
Key observation is that the fundamental representation ${\bf 6}$ of $SU(6)$ 
contains a doublet of $SU(2)_L$ and 
symmetric products of ${\bf 6}$ easily introduces a triplet of $SU(2)_{L}$, 
which is exotic. 
We therefore focus on the possibility of 
totally antisymmetric tensor representations of $SU(6)$. 
We find that the minimal set to accommodate one generation 
of quarks and leptons contains  two ${\bf 6^*}$ and one ${\bf 20}$ 
representations. 
Their parity assignments are fixed according to (\ref{parity}),   
\bea
{\bf 6^*} &=& \left\{ 
\begin{array}{l}
{\bf 6^*}_L = \underbrace{({\bf 3^*,1})_{(1/3,-1)}^{(+,-)} \oplus 
l_L({\bf 1,2})_{(-1/2,-1)}^{(+,+)}}_{{\bf 5^*}} 
\oplus \underbrace{({\bf 1,1})_{(0,5)}^{(-,-)}}_{{\bf 1}} \\
{\bf 6^*}_R = \underbrace{({\bf 3^*,1})_{(1/3,-1)}^{(-,+)} \oplus 
({\bf 1,2})_{(-1/2,-1)}^{(-,-)}}_{{\bf 5^*}} 
\oplus \underbrace{\nu_R({\bf 1,1})_{(0,5)}^{(+,+)}}_{{\bf 1}}  
\end{array}
\right. 
\label{12} \\
{\bf 6^*} &=& 
\left\{ 
\begin{array}{l}
{\bf 6^*}_L = \underbrace{({\bf 3^*,1})_{(1/3,-1)}^{(-,-)} \oplus 
({\bf 1,2})_{(-1/2,-1)}^{(-,+)}}_{{\bf 5^*}} \oplus 
\underbrace{({\bf 1,1})_{(0,5)}^{(+,-)}}_{{\bf 1}} \\
{\bf 6^*}_R = \underbrace{d_R^*({\bf 3^*,1})_{(1/3,-1)}^{(+,+)} \oplus 
({\bf 1,2})_{(-1/2,-1)}^{(+,-)}}_{{\bf 5^*}} \oplus 
\underbrace{({\bf 1,1})_{(0,5)}^{(-,+)}}_{{\bf 1}} \\
\end{array}
\right. 
\\
{\bf 20} &=& \left\{
\begin{array}{l}
{\bf 20}_L = \underbrace{q_L({\bf 3,2})_{(1/6,-3)}^{(+,+)} \oplus 
({\bf 3^*,1})_{(-2/3,-3)}^{(+,-)} 
\oplus ({\bf 1,1})_{(1,-3)}^{(+,-)}}_{{\bf 10}} \\
\hspace*{15mm}\oplus \underbrace{({\bf 3^*,2})_{-1/6,3}^{(-,+)} \oplus 
({\bf 3,1})_{(2/3,-3)}^{(-,-)} \oplus 
({\bf 1,1})_{(-1,3)}^{(-,-)}}_{{\bf 10^*}} \\
{\bf 20}_R = \underbrace{({\bf 3,2})_{(1/6,-3)}^{(-,-)} \oplus 
({\bf 3^*,1})_{(-2/3,-3)}^{(-,+)} \oplus ({\bf 1,1})_{(1,-3)}^{(-,+)}}_{{\bf 10}} 
\\
\hspace*{15mm}\oplus \underbrace{({\bf 3^*,2})_{-1/6,3}^{(+,-)} \oplus 
u_R({\bf 3,1})_{(2/3,-3)}^{(+,+)} \oplus 
e_R({\bf 1,1})_{(-1,3)}^{(+,+)}}_{{\bf 10^*}} \\
\end{array}
\right. 
\label{fermionrepr}
\eea
where the numbers written by the bold face in the parenthesis are 
the representations under $SU(3)_C \times SU(2)_L$. 
The numbers written in the subscript denote the charges 
under $U(1)_Y \times U(1)_X$. 
$L(R)$ means the left(right)-handed chirality. 
Note that the difference between the first and the second 
${\bf 6^*}$ representations lies in 
the relative sign of the parity at $y= 0$. 
The corresponding representations of $SU(5)$ are also displayed. 
It is interesting that the charged lepton doublet $l_L$ and 
the right-handed down quark singlet $d_R^*$ are separately embedded 
in different ${\bf 5^*}$ representations. 
Similarly, the quark doublet $q_L$ and 
the right-handed up quark $u_R$, electron $e_R$ are separately 
embedded in different ${\bf 10}$ representations.

A remarkable fact is that one generation of quarks and leptons 
(including $\nu_{eR}$) is elegantly embedded as the zero modes of the 
minimal representations, without introducing any exotic particles.   
Since the zero mode sector is nothing but the matter content 
of the standard model (including $\nu_{eR}$), 
we have no 4D gauge anomalies with respect to the standard model gauge group. 
As the wave functions of zero modes are $y$-independent 
and the non-zero KK modes are vector-like, 
there is no anomalies even in the 5D sense. 
As for the remaining $U(1)_X$, we can easily see that the symmetry is anomalous 
and is broken at the quantum level. 
Thus its gauge boson should become heavy and 
is expected to be decoupled from the low energy sector of the theory \cite{SSS}. 

Next, let us study whether we can obtain 
the correct pattern of electroweak symmetry breaking 
$SU(2)_L \times U(1)_Y \to U(1)_{em}$. 
We will see below that for such purpose 
the minimal set of matter fields is not sufficient 
and we need to introduce several massless fermions 
belonging to the adjoint representation of $SU(6)$.  

\begin{figure}[htb]
 \begin{center}
  \includegraphics[width=5.5cm]{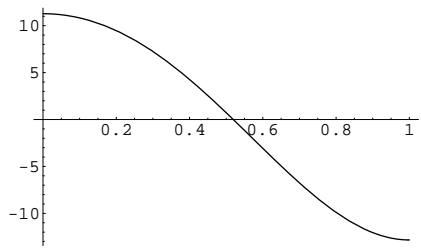}
 \end{center}
% \vspace*{-0.5cm}
\caption{One-loop Higgs potential with no adjoint fermion, 
massless fermions of $3 \times ({\bf 6^*} + {\bf 6^*})$ 
and $3 \times {\bf 20}$. The potential minimum is located at $\alpha=1$.} 
\label{noadj6620}
\end{figure}

One-loop induced Higgs ($A_5$) potential 
due to the matter fields 
$N_{{\rm ad}} \times {\bf 35} \oplus 3 \times (2 \times {\bf 6^*} 
\oplus {\bf 20})$ ($N_{{\rm ad}} $ and 3 denote the number of adjoint fermions 
and 3 generations, respectively) is calculated as 
\bea
V(\alpha) &=& C
\left[(4N_{{\rm ad}}-3)
\sum_{n=1}^\infty \frac{1}{n^5} 
\left\{ \cos(2\pi n \alpha)+2 \cos(\pi n \alpha)+6(-1)^n 
\cos(\pi n \alpha) \right\} \right. \nonumber \\
&& \left. + 48 \sum_{n=1}^\infty 
\frac{1+(-1)^n}{n^5} \cos(\pi n \alpha)
\right] 
\label{masslesshpotseveraladj6620}
\eea
where $C \equiv \frac{3}{128\pi^7 R^5}$.  
The dimensionless parameter $\alpha$ is defined by 
$\langle A_5 \rangle \equiv \frac{\alpha}{gR}\frac{\lambda_{27}}{2}$ 
where $\lambda_{27}$ is the twenty seventh generator of $SU(6)$ 
possessing the values in the (2,6) component of $6 \times 6$ matrix. 
As can be seen in Fig. \ref{noadj6620}, 
if we have no adjoint fermion $N_{{\rm ad}} = 0$, 
the potential is minimized at $\alpha=1$ 
where the desired electroweak symmetry breaking is not realized, 
namely $SU(2) \times U(1) \to U(1) \times U(1)$. 
%%%%%%%%%%%%%%%%%
%%%%%%%%%%%%%%%%%%%%
%%%%%%%%%%%%%%%%%
\begin{figure}[htb]
 \begin{center}
  \includegraphics[width=3.5cm]{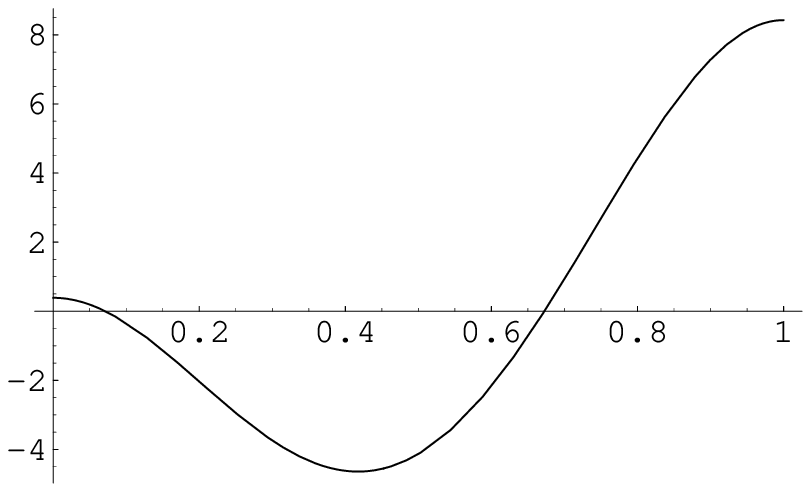}
  \hspace*{2mm}
  \includegraphics[width=3.5cm]{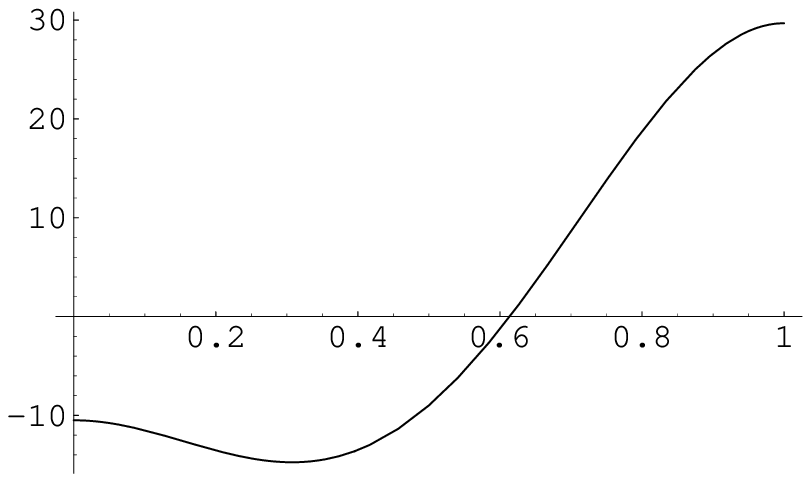}
%\\
\hspace*{2mm}
  \includegraphics[width=3.5cm]{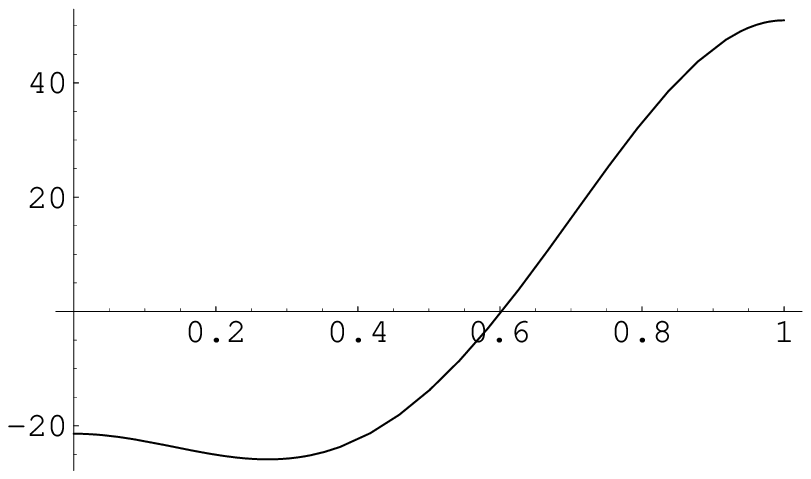}
  \hspace*{2mm}
  \includegraphics[width=3.5cm]{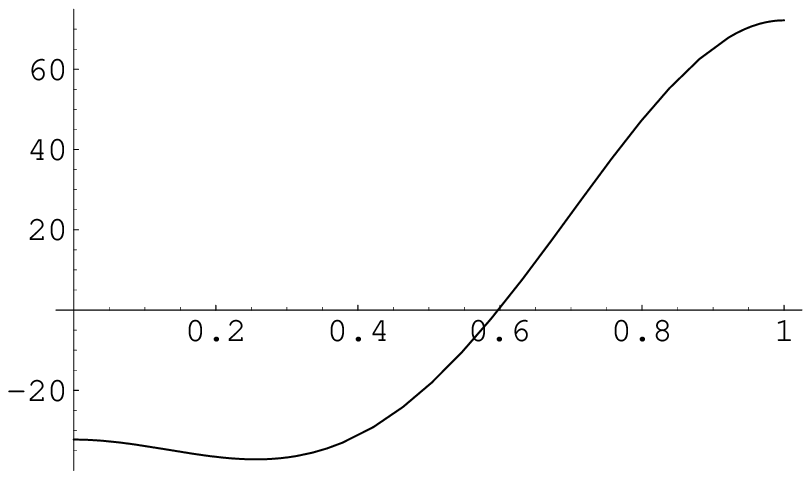}
 \end{center}
% \vspace*{-0.5cm}
\caption{One-loop Higgs potential with massless fermions of several adjoints, 
$3 \times({\bf 6^*} + {\bf 6^*} + {\bf 20})$ representations.
The plots from the left to the right correspond to 
the cases with one to four adjoint fermions. 
Their minimum is located at $\alpha = 0.417571, 0.307592, 0.27334, 0.256505$, 
respectively.} 
\label{severaladj6620}
\end{figure}
%%%%%%%%%%%%%%%%%%%%
On adding the adjoint fermions, 
we can see from Fig. \ref{severaladj6620} that the nontrivial minimum 
appears in the range $0 < \alpha < 1$ 
where the desired electroweak symmetry breaking 
$SU(2)_L \times U(1)_Y \to U(1)_{em}$ is realized. 
Note that the value of $\alpha$ at the minimum tends to become smaller, 
as the number of the adjoint fermions is larger.  
This feature is useful in order to make a Higgs mass heavy.   
Higgs mass can be obtained from the second derivative of the potential as
\bea
m_H^2 &=& \left. 
g^2 R^2 \frac{d^2V(\alpha)}{d \alpha^2} 
\right|_{\alpha = \alpha_0} \nonumber \\
%&=& 
%- \frac{3g_4^2}{32\pi^4 R^2} \left. 
%\left[(4N_{{\rm ad}}-3)
%\sum_{n=1}^\infty \frac{1}{n^3}
%\left\{
%2\cos(2\pi n \alpha) + \cos(\pi n \alpha) + 3(-1)^n \cos(\pi n \alpha)
%\right\} \right. \right. \nonumber \\
%&& \left. \left. 
%+24 \sum_{n=1}^\infty \frac{1+(-1)^n}{n^3}\cos(\pi n \alpha)
%\right] \right|_{\alpha=\alpha_0} \\
&=& \left. 
- \frac{3g_4^2 M_W^2}{32\pi^4 \alpha^2}
\left[(4N_{{\rm ad}}-3)
\sum_{n=1}^\infty \frac{1}{n^3}
\left\{
2\cos(2\pi n \alpha) + \cos(\pi n \alpha) + 3(-1)^n \cos(\pi n \alpha)
\right\} \right. \right. \nonumber \\
&& \left. \left. 
+24 \sum_{n=1}^\infty \frac{1+(-1)^n}{n^3}\cos(\pi n \alpha)
\right] \right|_{\alpha=\alpha_0}
\label{hmass6620}
\eea
where $\alpha_0$ denotes the value of $\alpha$ at the minimum of 
the potential. 
The relation derived from the gauge-Higgs unification $M_W = \alpha_0/R$ 
is used in the last expression. 
The gauge coupling in four dimensions $g_4$ is related to 
the gauge coupling in five dimensions $g$ through $g_4^2 = \frac{g^2}{2\pi R}$. 
The Higgs masses for several choices of $N_{{\rm ad}}$ are numerically 
calculated and tabulated in the table below. 
\begin{center}
\begin{tabular}{ccc}
\hline
Adj No. & $\alpha_0$ & Higgs mass \\
\hline
20 & 0.216557 & 113.9 $g_4$ GeV \\
21 & 0.216083 & 116.9 $g_4$ GeV \\
\hline
\end{tabular}
\end{center}
We can obtain a viable Higgs mass 
satisfying the experimental lower bound 
if more than 20 adjoint fermions are introduced 
($g_4$ is assumed to be ${\cal O}(1)$). 
Here we note that this result is just an existence proof 
not a realistic example for getting a relatively heavy Higgs mass. 
In our results, the compactification scale is a little bit low, 
$1/R = M_W/\alpha_0 \simeq 370$ GeV, 
which contradics with the current experimental data. 
Therefore, we need further investigations for obtaining a viable Higgs mass 
with more realsitic situation. 
As one of the possibilities, 
it would be interesting to analyze the Higgs potential 
with only three pairs of fermions in the representations 
$({\bf 6}^* + {\bf 6}^* + {\bf 20})$ on the warped space 
since it has been suggested that the Higgs mass on the warped space 
is enhanced comparing to the case of flat space \cite{HM}.

The extension to the case of massive fermion is straightforward. 
We can incorporate a $Z_{2}$-odd bulk mass of the type 
$M \epsilon (y) \ (\epsilon (y): \mbox{sign function})$ for fermions. 
The Higgs potential is then given by \cite{MT2}
\bea
V(\alpha) &=& C
\left[(4N_{{\rm ad}}-3)  
\sum_{n=1}^\infty \frac{1}{n^5} 
\left(1 + nz + \frac{1}{3}n^2 z^2 \right)e^{-nz} \right. \nonumber \\
&& \left. \times 
\left\{ \cos(2\pi n \alpha)+2 \cos(\pi n \alpha)+6(-1)^n 
\cos(\pi n \alpha) \right\} \right. \nonumber \\
&& \left. + 48\sum_{n=1}^\infty 
\left(1 + nz + \frac{1}{3}n^2 z^2 \right)e^{-nz}
\frac{1+(-1)^n}{n^5} \cos(\pi n \alpha) 
%&& \left. + 36 \sum_{n=1}^\infty 
%\left(1 + nz + \frac{1}{3}n^2 z^2 \right)e^{-nz}
%\frac{1+(-1)^n}{n^5}\cos(\pi n \alpha)
\right] 
\eea
where $z \equiv 2\pi R M$ and the bulk masses of fermions are taken to 
be a common value $M$ for simplicity. 
We will skip all the detail of the analysis by use of this potential, 
except reporting that there do not appear any drastic 
qualitative and quantitative change from the case of $M = 0$.

The relation $M_W = \alpha_0/R$ tells us that 
the compactification scale $1/R$ is not so far from the weak scale $M_{W}$, 
and therefore the GUT scale also cannot be extremely greater 
than the weak scale $M_{W}$ 
(Above the compactification scale, 
a power-law running of gauge couplings is expected  \cite{DDG}). 
Thus we have to worry about possible too rapid proton decay. 
Interestingly, such baryon number violating amplitude 
concerning KK zero modes turns out to be forbidden at least at the tree level. 
This is essentially because the quarks and leptons 
are separated into different representations in our model 
although they are accommodated in the same representation 
in ordinary $SU(5)$ GUT. 
From (\ref{fermionrepr}) we learn the type of baryon number (and lepton number) 
changing vertices of $A_{\mu}$ and $A_{5}$ is limited. 
Namely concerning fermionic zero-modes, only possibility is 
$u_{R} \leftrightarrow e_{R}$ due to ``colored" 4D gauge boson 
$A_{\mu}$ with ($SU(3)$, $SU(2)$) quantum number $({\bf 3, 1})$ 
and $q_{L} \leftrightarrow e_{R}$ due to  ``lepto-quark" 4D scalar $A_5$ 
with $({\bf 3,2})$. 
Let us note these relevant $A_{\mu}$ and $A_5$ 
are ``bosonic partners" of colored Higgs and $X, Y$ gauge bosons 
in ordinary $SU(5)$ GUT. 
Though these bosons couple to baryon number changing currents, 
these interactions do not lead to net baryon number violation, 
since each of gauge or Higgs boson couples to 
unique baryon number violating current. 
In the diagrams where these bosons are exchanged, 
one vertex with $\Delta B = 1/3$ and another vertex with $\Delta B = -1/3$ 
which is the Hermitian conjugate of the other necessarily appear, 
thus leading to no net baryon number violation. 
In other words, we can assign definite baryon (and lepton) number to each boson, and in such a sense baryon number is preserved at each vertex. 
It will be definitely necessary to consider 
whether such mechanism to preserve net baryon number is also operative 
at loop diagrams, though it is not discussed here.   

In summary, we have investigated a %realistic 
5D $SU(6)$ grand gauge-Higgs unification model 
compactified on an orbifold $S^1/Z_2$, with a realistic matter content. 
Three generation of quarks and leptons with additional right-handed neutrinos 
are just embedded as the zero modes of the minimal set of representations, 
$3 \times (2 \times {\bf 6^*} + {\bf 20})$, 
without introducing any exotic particles in the same representations. 
As a remarkable feature of the model, 
we have found the dangerous proton decay is forbidden at the tree level. 
We have also found that the desired pattern of electroweak symmetry breaking 
%$SU(2)_L \times U(1)_Y \to U(1)_{em}$ 
is dynamically realized by introducing suitable number of adjoint fermions. 
Higgs mass was also calculated and shown to become heavy, 
if certain number of the adjoint fermions are introduced. 
Searching for a simpler matter content yielding a reasonable Higgs mass 
is desirable and a very nontrivial task. 
This is left for a future work.

There are still many issues to be studied. 
%As we have briefly discussed above, 
The construction of the realistic hierarchy of Yukawa couplings 
is a fundamental problem in the gauge-Higgs unification 
since Yukawa coupling is naively the gauge coupling which is flavor independent. 
One of the promising proposals to avoid the problem \cite{CGM} 
is that the mixing between the bulk massive fermions and 
the brane localized quarks and leptons 
generates non-local Yukawa couplings 
after integrating out the bulk massive fermions. 
The huge hierarchy is then realized by an order one tuning of the bulk mass. 
It is very important to study whether this proposal can be incorporated 
into the present model. 
%To determine the GUT scale, or equivalently 
To study the energy evolution of gauge couplings and their unification 
is another important issue. 
These issues will be discussed elsewhere.

\begin{center}
{\bf Acknowledgments}
\end{center}
The author would like to thank the organizers of the workshop 
for providing me with an oppotunity to talk at such a nice workshop. 
The work of the author was supported 
in part by the Grant-in-Aid for Scientific Research 
of the Ministry of Education, Science and Culture, No.18204024.
%\end{theacknowledgments}



\bibliographystyle{aipproc}   % if natbib is available

\bibliography{sample}

\begin{thebibliography}{9}

\bibitem{LM2}
 C.S.~Lim and N.~Maru, \emph{Phys. Lett.} \textbf{B653}, 320 (2007). 


\bibitem{Manton} 
  N.~S.~Manton,
  %``A New Six-Dimensional Approach To The Weinberg-Salam Model,''
  \emph{Nucl. Phys.}  \textbf{B158}, 141 (1979).

\bibitem{Fairlie}
  D.~B.~Fairlie,
  %``Higgs' Fields And The Determination Of The Weinberg Angle,''
  \emph{Phys. Lett.} \textbf{B82}, 97 (1979); 
  %``Two Consistent Calculations Of The Weinberg Angle,''
  \emph{J. Phys.} \textbf{G 5}, L55 (1979).


%%%%%%%%%%%%%%%%%%%%%%%%%%%%%%%
\bibitem{Hosotani}  
  Y.~Hosotani,
    %``Dynamical Mass Generation By Compact Extra Dimensions,''
  \emph{Phys. Lett.} \textbf{B126}, 309 (1983); 
  %``Dynamical Gauge Symmetry Breaking As The Casimir Effect,''
  \emph{Phys. Lett.} \textbf{B129}, 193 (1983); 
  %``DYNAMICS OF NONINTEGRABLE PHASES AND GAUGE SYMMETRY BREAKING,''
  \emph{Annals Phys.} \textbf{190}, 233 (1989).

%%%%%%%%%%%%%%% 

%%%%%%%%%%%%%%%%%%%%%%%%%%%%%%%%%%%%%%%%%%%%%%%%%%%%%%%%%%%%%%% 
\bibitem{HIL}
  H.~Hatanaka, T.~Inami and C.~S.~Lim,
  %``The gauge hierarchy problem and higher dimensional gauge theories,''
  \emph{Mod. Phys. Lett.} \textbf{A13}, 2601 (1998); 
%  [arXiv:hep-th/9805067].
%%%%%%%%%%%%%%%%%%%%%%%%%%%%%%%%%%%%%%%%%%%%%%%%%%%%%%%%%%%%%%
\bibitem{ABQ}
  I.~Antoniadis, K.~Benakli and M.~Quiros,
  %``Finite Higgs mass without supersymmetry,''
  \emph{New J. Phys.} \textbf{3}, 20 (2001). 
%  [arXiv:hep-th/0108005].


\bibitem{GIQ}
  G.~von Gersdorff, N.~Irges and M.~Quiros,
  %``Bulk and brane radiative effects in gauge theories on orbifolds,''
  \emph{Nucl. Phys.} \textbf{B635}, 127 (2002).

\bibitem{CNP}
  The first reference in \cite{ACP}. 

\bibitem{HLM}
  K.~Hasegawa, C.~S.~Lim and N.~Maru,
  %``An attempt to solve the hierarchy problem based on gravity gauge Higgs
  %unification scenario,''
  \emph{Phys. Lett.}\textbf{B604}, 133 (2004). 
%  [arXiv:hep-ph/0408028].



\bibitem{LMH}
  C.~S.~Lim, N.~Maru and K.~Hasegawa,
  %``Six dimensional gauge-Higgs unification with an extra space S**2 and the
  %hierarchy problem,''
  arXiv:hep-th/0605180.


\bibitem{MY}
  N.~Maru and T.~Yamashita,
  %``Two-loop calculation of Higgs mass in gauge-Higgs unification: 
  %5D  massless QED compactified on S**1,''
  \emph{Nucl. Phys.} \textbf{B754}, 127 (2006); 
%  [arXiv:hep-ph/0603237].
  Y.~Hosotani, N.~Maru, K.~Takenaga and T.~Yamashita,
  %``Two loop finiteness of Higgs mass and potential in the gauge-Higgs
  %unification,''
\emph{Prog. Theor. Phys.} \textbf{118}, 1053 (2007). 



\bibitem{LM}
  C.~S.~Lim and N.~Maru,
  %``Calculable one-loop contributions to S and T parameters in the gauge-Higgs
  %unification,''
    \emph{Phys. Rev.} \textbf{D75}, 115011 (2007). 

\bibitem{ALM}
  Y.~Adachi, C.~S.~Lim and N.~Maru,
  %``Finite Anomalous Magnetic Moment in the Gauge-Higgs Unification,''
  \emph{Phys. Rev.} \textbf{D76}, 075009 (2007)

\bibitem{Falkowski}
  A.~Falkowski,
  %``Pseudo-Goldstone Higgs Production via Gluon Fusion,''
  arXiv:0711.0828 [hep-ph].
  
\bibitem{MO}
  N.~Maru and N.~Okada,
  %``Gauge-Higgs Unification at LHC,''
  arXiv:0711.2589 [hep-ph](To appear in PRD).

\bibitem{LR} 
 I.~Low, talk at E\"ot\"ovs-Cornell 2007 Summer Workshop on Particle Theory, 
 {\bf http://www.lns.cornell.edu/\~{}eotcor07/budapest\_ianlow.pdf}; 
 I.~Low and R.~Rattazzi, to appear. 

\bibitem{KLY}
  M.~Kubo, C.~S.~Lim and H.~Yamashita,
%  ``The Hosotani mechanism in bulk gauge theories with an orbifold 
%  extra space S(1)/Z(2),''
  \emph{Mod. Phys. Lett.} \textbf{A17}, 2249 (2002). 

\bibitem{CGM}
  C.~Cs\'aki, C.~Grojean and H.~Murayama,
  %``Standard model Higgs from higher dimensional gauge fields,''
  \emph{Phys. Rev.} \textbf{D67}, 085012 (2003). 
%  [arXiv:hep-ph/0210133].


\bibitem{SSS}
C.~A.~Scrucca, M.~Serone and L.~Silvestrini,
  %``Electroweak symmetry breaking and fermion masses from extra dimensions,''
  \emph{Nucl. Phys.} \textbf{B669}, 128 (2003). 


\bibitem{GMN}
  I.~Gogoladze, Y.~Mimura and S.~Nandi,
  %``Gauge Higgs unification on the left-right model,''
  \emph{Phys. Lett.} \textbf{B560}, 204 (2003); 
%  [arXiv:hep-ph/0301014]; 
  %``Coupling unifications in gauge-Higgs unified orbifold models,''
  \emph{Phys. Rev.} \textbf{D72}, 055006 (2005).  
  %[arXiv:hep-ph/0504082], 

\bibitem{HHKY}
  N.~Haba, Y.~Hosotani, Y.~Kawamura and T.~Yamashita,
  %``Dynamical symmetry breaking in gauge-Higgs unification on orbifold,''
  \emph{Phys. Rev.} \textbf{D70}, 015010 (2004). 
  %[arXiv:hep-ph/0401183].

\bibitem{HNT}
Y.~Hosotani, S.~Noda and K.~Takenaga,
  %``Dynamical gauge symmetry breaking and mass generation on the orbifold
  %T**2/Z(2),''
  \emph{Phys. Rev.} \textbf{D69}, 125014 (2004); 
%  [arXiv:hep-ph/0403106]
  %``Dynamical gauge-Higgs unification in the electroweak theory,''
  \emph{Phys.\ Lett.} \textbf{B607}, 276 (2005).  
%  [arXiv:hep-ph/0410193].
  
  
\bibitem{BQ}
  C.~Biggio and M.~Quiros,
  %``Higgs-gauge unification without tadpoles,''
  \emph{Nucl. Phys.} \textbf{B703}, 199 (2004). 
%  [arXiv:hep-ph/0407348].

\bibitem{MSSS}
G.~Martinelli, M.~Salvatori, C.~A.~Scrucca and L.~Silvestrini,
  %``Minimal gauge-Higgs unification with a flavour symmetry,''
  \emph{JHEP} \textbf{0510}, 037 (2005). % [arXiv:hep-ph/0503179]. 


%%%%%%%%%%%%%%% 


\bibitem{OW}
  K.~y.~Oda and A.~Weiler,
  %``Wilson lines in warped space: Dynamical symmetry breaking and
  %restoration,''
  \emph{Phys. Lett.} \textbf{B606}, 408 (2005).  
  %[arXiv:hep-ph/0410061]. 

\bibitem{ACP}
 R.~Contino, Y.~Nomura and A.~Pomarol,
  %``Higgs as a holographic pseudo-Goldstone boson,''
  \emph{Nucl. Phys.} \textbf{B671}, 148 (2003); 
%  [arXiv:hep-ph/0306259].

K.~Agashe, R.~Contino and A.~Pomarol,
  %``The minimal composite Higgs model,''
  \emph{Nucl. Phys.} \textbf{B719}, 165 (2005);  
%  [arXiv:hep-ph/0412089]; 
  K.~Agashe and R.~Contino,
  %``The minimal composite Higgs model and electroweak precision tests,''
  \emph{Nucl.\ Phys.} \textbf{B742}, 59 (2006); 
%  [arXiv:hep-ph/0510164].
  K.~Agashe, R.~Contino, L.~Da Rold and A.~Pomarol,
  %``A custodial symmetry for Z b anti-b,''
  \emph{Phys. Lett.} \textbf{B641}, 62 (2006); 
%  [arXiv:hep-ph/0605341].
  R.~Contino, L.~Da Rold and A.~Pomarol,
  %``Light custodians in natural composite Higgs models,''
    \emph{Phys. Rev.} \textbf{D75}, 055014 (2007).


\bibitem{HM}
 Y.~Hosotani and M.~Mabe,
  %``Higgs boson mass and electroweak-gravity hierarchy from dynamical
  %gauge-Higgs unification in the warped spacetime,''
  \emph{Phys. Lett.} \textbf{B615}, 257 (2005); 
  %[arXiv:hep-ph/0503020]. 
Y.~Hosotani, S.~Noda, Y.~Sakamura and S.~Shimasaki,
  %``Gauge-Higgs unification and quark-lepton phenomenology in the warped
  %spacetime,'' 
  \emph{Phys. Rev.} \textbf{D73}, 096006 (2006); 
  %[arXiv:hep-ph/0601241]. 
  Y.~Sakamura and Y.~Hosotani,
  %``W W Z, W W H, and Z Z H couplings in the dynamical gauge-Higgs  
  %unification in the warped spacetime,''
  \emph{Phys. Lett.} \textbf{B645}, 442 (2007); 
%  arXiv:hep-ph/0607236; 
  Y.~Hosotani and Y.~Sakamura,
  %``Anomalous Higgs couplings in the SO(5) x U(1)(B-L) gauge-Higgs unification
  %in warped spacetime,''
   \emph{Prog. Theor. Phys.} \textbf{118}, 935 (2007);
   Y.~Sakamura,
  %``Effective theories of gauge-Higgs unification models in warped spacetime,''
    \emph{Phys. Rev.} \textbf{D76}, 065002 (2007).

\bibitem{CPSW}
  M.~S.~Carena, E.~Ponton, J.~Santiago and C.~E.~M.~Wagner,
  %``Electroweak constraints on warped models with custodial symmetry,''
  \emph{Phys. Rev.} \textbf{D76}, 035006 (2007); 
%  [arXiv:hep-ph/0701055]. 
  A.~D.~Medina, N.~R.~Shah and C.~E.~M.~Wagner,
  %``Gauge-Higgs Unification and Radiative Electroweak Symmetry Breaking in
  %Warped Extra Dimensions,''
   \emph{Phys.\ Rev.} \textbf{D76}, 095010 (2007);
 M.~Carena, A.~D.~Medina, B.~Panes, N.~R.~Shah and C.~E.~M.~Wagner,
  %``Collider Phenomenology of Gauge-Higgs Unification Scenarios in Warped Extra
  %Dimensions,''
  arXiv:0712.0095 [hep-ph].


\bibitem{PW}
  G.~Panico and A.~Wulzer,
  %``Effective Action and Holography in 5D Gauge Theories,''
   \emph{JHEP} \textbf{0705}, 060 (2007). 
%  arXiv:hep-th/0703287.



\bibitem{PS}
G.~Panico and M.~Serone,
  %``The electroweak phase transition on orbifolds with gauge-Higgs
  %unification,''
  \emph{JHEP} \textbf{0505}, 024 (2005); %[arXiv:hep-ph/0502255]. 
G.~Panico, M.~Serone and A.~Wulzer,
  %``A model of electroweak symmetry breaking from a fifth dimension,''
  \emph{Nucl. Phys.} \textbf{B739}, 186 (2006); %arXiv:hep-ph/0510373; 
  %``Electroweak symmetry breaking and precision tests with a fifth
  %dimension,''
  \emph{Nucl. Phys.} \textbf{B762}, 189 (2007). 
%  [arXiv:hep-ph/0605292].



\bibitem{CCP}
G.~Cacciapaglia, C.~Csaki and S.~C.~Park,
  %``Fully radiative electroweak symmetry breaking,''
  \emph{JHEP} \textbf{0603}, 099 (2006); %arXiv:hep-ph/0510366. 

\bibitem{HMOY}
 N.~Haba, S.~Matsumoto, N.~Okada and T.~Yamashita,
  %``Effective theoretical approach of gauge-Higgs unification model and its
  %phenomenological applications,''
  \emph{JHEP} \textbf{0602}, 073 (2006);
  %[hep-ph/0511046]
  I.~Gogoladze, N.~Okada and Q.~Shafi,
  %``Higgs Boson Mass From Gauge-Higgs Unification,''
  \emph{Phys. Lett.} \textbf{B655}, 257 (2007); 
  \emph{Phys. Lett.} \textbf{B659}, 316 (2008). 


\bibitem{MT}
  N.~Maru and K.~Takenaga,
  %``Aspects of phase transition in gauge-Higgs unification at finite
  %temperature,''
  \emph{Phys. Rev.} \textbf{72}, 046003 (2005);   
%  [arXiv:hep-th/0505066]; 
  %``Effects of bulk mass in gauge-Higgs unification,''
  \emph{Phys. Lett.} \textbf{B637}, 287 (2006); 
%  [arXiv:hep-ph/0602149].   
  %``Bulk mass effects in gauge-Higgs unification at finite temperature,''
  \emph{Phys. Rev.} \textbf{D74}, 015017 (2006). 
%  [arXiv:hep-ph/0606139]. 

%%%%% 
\bibitem{Hosotani2}
  Y.~Hosotani,
  %``All-order finiteness of the Higgs boson mass in the dynamical gauge-Higgs
  %unification,''
  arXiv:hep-ph/0607064. 

\bibitem{ST}
  M.~Sakamoto and K.~Takenaga,
  %``Large gauge hierarchy in gauge-Higgs unification,''
  \emph{Phys. Rev.} \textbf{D75}, 045015 (2007). 
%  [arXiv:hep-th/0609067].

\bibitem{IKT}
  K.~Inoue, A.~Kakuto and H.~Takano,
%  ``Higgs As (Pseudo)Goldstone Particles,''
  \emph{Prog. Theor. Phys.} \textbf{75}, 664 (1986).

\bibitem{AJ}
  A.~A.~Anselm and A.~A.~Johansen,
  %``Susy GUT With Automatic Doublet - Triplet Hierarchy,''
  \emph{Phys.\ Lett.} \textbf{B200}, 331 (1988);  
  A.~A.~Anselm,
  %``A SUPERSYMMETRIC THEORY OF GRAND UNIFICATION WITH AUTOMATIC HIERARCHY AND
  %LOW-ENERGY PHYSICS,''
  \emph{Sov. Phys. JETP} \textbf{67} (1988) 663
  [\emph{Zh. Eksp. Teor. Fiz.}\textbf{94} (1988) 26].

\bibitem{BD}
  Z.~G.~Berezhiani and G.~R.~Dvali,
  %``Possible Solution Of The Hierarchy Problem In Supersymmetrical Grand
  %Unification Theories,''
  \emph{Bull. Lebedev Phys. Inst.} \textbf{5} (1989) 55
  [\emph{Kratk. Soobshch. Fiz.} \textbf{5} (1989) 42]. 

\bibitem{BDS}
  R.~Barbieri, G.~R.~Dvali and A.~Strumia,
  %``Grand Unified Supersymmetric Higgs Bosons As Pseudogoldstone Particles,''
  \emph{Nucl. Phys.} \textbf{B391}, 487 (1993); 
%
%\bibitem{BDM}
  R.~Barbieri, G.~R.~Dvali and M.~Moretti,
  %``Back to the doublet - triplet splitting problem,''
  \emph{Phys. Lett.} \textbf{B312}, 137 (1993).


\bibitem{BDSBH}
  R.~Barbieri, G.~R.~Dvali, A.~Strumia, Z.~Berezhiani and L.~J.~Hall,
  %``Flavor in supersymmetric grand unification: A Democratic approach,''
  \emph{Nucl. Phys.} \textbf{B432}, 49 (1994). 
%  [arXiv:hep-ph/9405428].

\bibitem{BCR}
  Z.~Berezhiani, C.~Csaki and L.~Randall,
  %``Could the supersymmetric Higgs particles naturally be pseudoGoldstone
  %bosons?,''
  \emph{Nucl. Phys.} \textbf{B444}, 61 (1995). 
%  [arXiv:hep-ph/9501336].

\bibitem{Berezhiani}
  Z.~Berezhiani,
  %``Susy SU(6) Gift For Doublet-Triplet Splitting And Fermion Masses,''
  \emph{Phys. Lett.} \textbf{B355}, 481 (1995). 
%  [arXiv:hep-ph/9503366].

\bibitem{ACG}  
 N.~Arkani-Hamed, A.G.~Cohen and H.~Georgi, 
  \emph{Phys. Lett.} \textbf{B513}, 232 (2001); 
 N.~Arkani-Hamed, A.G.~Cohen, E. Katz and A.E.~Nelson, 
 \emph{JHEP} \textbf{0207}, 034 (2002); 
 M.~Schmaltz and D. Tucker-Smith, 
 \emph{Ann. Rev. Nucl. Part. Sci.}\textbf{55} 229 (2005); 
 M.~Perelstein, hep-ph/0512128.  
 
\bibitem{holography}
  J.~M.~Maldacena,
  %``The large N limit of superconformal field theories and supergravity,''
  \emph{Adv. Theor. Math. Phys.}\textbf{2}, 231 (1998)
  [\emph{Int. J. Theor. Phys.} \textbf{38}, 1113 (1999)]; 
%  [arXiv:hep-th/9711200].
  S.~S.~Gubser, I.~R.~Klebanov and A.~M.~Polyakov,
  %``Gauge theory correlators from non-critical string theory,''
  \emph{Phys. Lett.} \textbf{B428}, 105 (1998); 
%  [arXiv:hep-th/9802109].
  E.~Witten,
  %``Anti-de Sitter space and holography,''
  \emph{Adv. Theor. Math. Phys.} \textbf{2}, 253 (1998). 
%  [arXiv:hep-th/9802150].

\bibitem{holography2}
  N.~Arkani-Hamed, M.~Porrati and L.~Randall,
  %``Holography and phenomenology,''
  \emph{JHEP} \textbf{0108}, 017 (2001);
  R.~Rattazzi and A.~Zaffaroni,
  %``Comments on the holographic picture of the Randall-Sundrum model,''
  \emph{JHEP} \textbf{0104}, 021 (2001). 


\bibitem{BN}
  G.~Burdman and Y.~Nomura,
%  ``Unification of Higgs and gauge fields in five dimensions,''
  \emph{Nucl. Phys.} \textbf{B656}, 3 (2003). 
%  [arXiv:hep-ph/0210257].

\bibitem{Kawamura} 
   Y.~Kawamura,
  %``Triplet-doublet splitting, proton stability and extra dimension,''
  \emph{Prog. Theor. Phys.} \textbf{105}, 999 (2001); 
%  [arXiv:hep-ph/0012125].
%``Split multiplets, coupling unification and extra dimension,''
  \emph{Prog. Theor. Phys.} \textbf{105}, 691 (2001). 
%  [arXiv:hep-ph/0012352].


\bibitem{MT2}
The second reference in \cite{MT}. 

\bibitem{DDG}
  K.~R.~Dienes, E.~Dudas and T.~Gherghetta,
  %``Extra spacetime dimensions and unification,''
  \emph{Phys.\ Lett.} \textbf{B436}, 55 (1998). 
%  [arXiv:hep-ph/9803466].



\end{thebibliography}

\IfFileExists{\jobname.bbl}{}
 {\typeout{}
  \typeout{******************************************}
  \typeout{** Please run "bibtex \jobname" to optain}
  \typeout{** the bibliography and then re-run LaTeX}
  \typeout{** twice to fix the references!}
  \typeout{******************************************}
  \typeout{}
 }

\end{document}